\documentclass{amsart}
\usepackage{amssymb}
\usepackage{verbatim}
\newtheorem{Definition}{Definition}

\def \RR {{\bf R}}

\begin{document}
\begin{abstract}
{For solutions of (inviscid, forceless, one dimensional) Burgers
equation with random initial condition, it is heuristically shown 
 that a stationary
Feller-Markov property (with respect to
the space variable) at some time is conserved at later times, and an evolution
equation is derived for the infinitesimal generator. Previously known explicit
solutions such as Frachebourg-Martin's (white noise initial velocity) and
Carraro-Duchon's L\'evy process intrinsic-statistical solutions (including
Brownian initial velocity) are recovered as special cases.}
\end{abstract}

\date{\today}
\title[Markov solutions of Burgers Equation]{Markovian solutions of inviscid
Burgers equation}
\author[M-L. Chabanol]{Marie-Line Chabanol}
\address{Institut Fourier (Grenoble)\\UMR 5582 CNRS-Universit\'e Joseph
Fourier\\100, rue des Math\'ematiques, B.P. 74\\
38041 Saint-Martin d'H\`eres Cedex}
\email{Marie-Line.Chabanol@ujf-grenoble.fr\\ Jean.Duchon@ujf-grenoble.fr}
\author[J. Duchon]{Jean Duchon}
\thanks{The authors thank Christophe Giraud for fruitful discussions, the
Erwin Schr\"o\-dinger Institute in Vienna for their invitation and support,
the organizers of the program ``developed turbulence'' there (Gaw\c edzki,
Kupiainen, Vergassola) and also Uriel Frisch and Yakov Sinai who conducted
the workshop ``Burgers turbulence and beyond''}
\maketitle
\section{Introduction}

We consider the inviscid Burgers equation
$ \partial_t u + \partial_x(\frac{1}{2} u^2) = 0$
with random initial data $u_0$. Burgers equation has originally been
introduced~\cite{Bur} as a 1D model of turbulence. Although it is now
clear that it does not exhibit lots of features of ``true''
turbulence, we nevertheless still think it is a good equation on which
one can try and find new methods to apply on Euler equation. Having
this in mind, taking random initial data seems quite a natural
problem. It is also physically relevant in the contexts of interface
dynamics, of aggregation of particles \cite{FM2}, and some others.
Burgers equation with a random force on the r.h.s.  has also been studied,
mainly as a ``benchmark'' to test methods designed for (Navier-Stokes)
forced turbulence, many of which turn out to produce spurious predictions
when applied to the simpler Burgers case.
See~\cite{EvdE} for
instance.

The case of a Brownian initial data has already been investigated by
Sinai~\cite{Sin}. Carraro and Duchon
\cite{CD2,CD} 
showed that L\'evy processes are conserved by Burgers
equation. They also obtained the explicit evolution equation for the
characteristic function of the L\'evy process solutions of
Burgers. A noticeable point is that they made no use of the Hopf-Cole
construction of the solution
(Bertoin~\cite{Ber} recovered essentially the same result with Hopf-Cole).

We will essentially follow Carraro and
Duchon :
first we define what we call a statistical solution of
Burgers equation, and write an infinite set of equations for
the $n$-point functions of such solutions. 
We show that the assumption that the process is
Feller (in space) for all time 
 yields an evolution equation for
the infinitesimal generator of this process. Conversely, a Feller process 
whose generator satisfies
this equation is a statistical solution of Burgers equation.
 This will allow us to recover Carraro and
Duchon's result on L\'evy processes, as a special case. Frachebourg and
Martin's explicit solution \cite{FM} in the case of an initial white noise
velocity is also a particular solution to our equation. 

\section{Notations and definitions}

A Markov process $u(x)_{x \in \RR}$ 
 can be characterized by its one point and its
transition  
probabilities $p_x(du)$ and $q_{x,y}(u,dv)$ , $x<y$, that satisfy,
$ \forall x_0<\cdots<x_k$ and $ f_i$ borelian positive,\qquad
$\mathbf{E}[\prod_{i=0}^k f_i(u(x_i))] =$  
\[ 
\int p_{x_0}(du_0)f_0(u_0) \int q_{x_0,x_1}(u_0,du_1) f_1(u_1) \cdots  \int
q_{x_{k-1},x_k}(u_{k-1},du_k) f_k(u_k).
\]

A Markov process is {\em homogeneous} if its
transition probabilities $q_{x,y}$ depend on $x$ and
$y$ only through $y-x$. In this case, we write $q_h$ instead of
$q_{x,x+h}$ . 

A process 
$u(x)_{x\in \RR}$ is {\em stationary} if and only if it is
translation invariant~: the law of $(u_{x+x_1}, \ldots, u_{x+x_n})$
does not depend on $x$. Hence a  Markov process is stationary if
and only if it is
homogeneous and its one point probability $p_x(du)$ does not depend on
$x$.

If $u$ is a homogeneous Markov process, $h>0$, and
$f$ is a continuous function vanishing at infinity, we put
 $Q_h f(u)\,=\,\int f(v)q_h(u,dv)$ .

 A {\em Feller} process is a homogeneous 
Markov process such that
 for each $f$, for each $h>0$,
  $Q_hf$ is also continuous and vanishes at infinity, and 
$\lim_{h\rightarrow 0} Q_h f= f$ pointwise. 

A Feller process always has a c\`adl\`ag version \cite{Yor}.

One can define the
{\em infinitesimal generator} of a Feller process~: it is the operator
$A$, 
defined for all the
functions $f$ such that the limit below exists,
by
\[
\forall \, x \in \RR, \, Af(x) = \lim_{h\rightarrow 0^+}\frac{Q_h f(x)  -
f(x)}{h}
\]

Formally, $Q_h\,=\,\exp (hA)$ , $Q'_h:=dQ/dh = A Q_h$ and an invariant
measure $p_0$ satisfies $\ ^t \!\! A p_0=0$ .


The
\section{Statistical solutions of Burgers equation}

We will closely follow~\cite{CD}  (see also~\cite{RRinFriedSerre}).
Let $E$ be the space of c\`adl\`ag real functions equipped with the
smallest $\sigma$-algebra  $\mathcal C(E)$
such that for each $x \in \RR$, $u \mapsto u(x)$ is measurable.
Let $\mathcal D$ be the set of real $C^\infty$ functions with compact
support.
A probability $\mu$ on $E$ is then characterized by its characteristic
function 
\[
 v \in {\mathcal D} \mapsto \int_E \exp \, [i \int_{\RR} u(x) v(x)
dx]\, d\mu(u)= \hat{\mu}(v).
\]
Let $u_0$~:  $(\Omega, {\mathcal A}, P) \to E$ be a random process, defined
on some probability space, and let $\mu_0 : \mathcal C(E)\to [0,1]$  denote 
its
probability law. Assume $u(x,t)$ is a (weak) solution of Burgers equation
with $u(.,0)=u_0$, $u(.,t) \in E$ for $t>0$, 
and everything makes sense in the following calculation ~: 
integrability, and differentiability with respect to $t$. 
Let $\mu_t$ denote the law of $u(.,t)$.
Formally, one then gets for each $v \in {\mathcal D}$~: 
\begin{eqnarray*}\label{stats}
\partial_t \hat{\mu}_t(v) &=& \int_E \partial_t \{\exp \, [ i \int_{\RR} u(x)
v(x)dx]\}\, d\mu_t(u)\\
&=& \int_E \partial_t \{\exp \, [i \int_{\RR} u(x,t) v(x)dx]\}\, d\mu_0(u_0)\\
&=& \int_E \exp \, [i \int_{\RR} u(x,t) v(x) dx)] \, \partial_t[i\int_{\RR}  
u(x,t) v(x) dx ]\,d\mu_0(u_0)\\
&=& \int_E  \exp \, [i \int_{\RR}u(x,t) v(x)dx]\,i \int_{\RR} \frac{1}{2} 
u(x,t)^2 v'(x) dx \,d\mu_0(u_0)\\
&=& i \int_E  \int_{\RR} \frac{1}{2} u(x)^2 v'(x) dx \, \exp \, [i \int_{\RR}
u\, v\,] \,d\mu_t(u)
\end{eqnarray*}

Hence our definition of a {\em statistical solution} of Burgers
equation~:

\begin{Definition}
A statistical solution of Burgers equation is a set $(\mu_t)_{t\geq
  0}$ of probabilities on $(E,\mathcal C(E))$ such that for any $v \in 
{\mathcal D}$,  
\begin{equation}  \partial_t \hat{\mu}_t(v) = i \int_E \int_{\RR}
\frac{1}{2} u(x)^2 v'(x) dx \exp \, [i \int_{\RR} u\, v\,]\, 
d\mu_t(u)\end{equation} \label{mut}
\end{Definition}

Let us assume now that we have a statistical solution of Burgers equation,
$(\mu_t)_{t\geq 0}$~, and that for all $t$, all the moments of $\mu_t$
are well defined. Then one can write 
\[
\exp \, [i \int_{\RR} u(x) v(x) dx] =
\sum_{n=0}^\infty \frac{i^n}{n!} \int_{\RR^n} \prod_{j=1}^{n} u(x_j)v(x_j)
dx_j
\]
 Equation (\ref{mut}) thus becomes $\forall \,v \in {\mathcal D}$,
\begin{multline} 2\sum_{n=0}^\infty \frac{i^n}{n!}
\int_{\RR^n} \partial_t {\bf E}[\prod_{j=1}^{n}u(x_j)v(x_j)] \prod dx_j 
=\\
 =i \sum_{n=0}^\infty \frac{i^n}{n!} \int_{\RR^{n+1}}  {\bf
E}[\prod_{j=1}^{n}u(x_j)v(x_j)u(x)^2 v'(x)] dx \prod dx_j  =\\
=i\sum_{n=0}^\infty i^n \int_{x_0<x_1<\ldots <x_n} \sum_{j=0}^{n}
{\bf E} [u(x_j)\frac{v'(x_j)}{v(x_j)}\prod_{k=0}^n u(x_k)v(x_k)]\prod dx_k
\label{evE}
\end{multline}

\section{Evolution equation for Markov solutions}

We are now looking for solutions such that at each time $t$, $x\mapsto 
u(x,t)$ is
a stationary Feller process (with respect to space $x$). We are going to show
that for such processes, the infinite set of equations (\ref{evE})
is equivalent to an evolution equation for the infinitesimal generator
of $u$.

We thus assume now that the solution 
$x \mapsto u(x,t)$ is a stationary Feller process, with one
point probability $p(du,t)$ and transition probability $q_h(u_1,du_2,t)$, the
equation (\ref{evE}) becomes $\forall \, v \in {\mathcal D}$~:


\begin{multline}
2\sum_{n=1}^{\infty} \int_{x_1<\dots<x_n}dx_1 \dots dx_n \ u_1v(x_1) \dots 
u_n v(x_n) \\
\times \partial_t [p(du_1) q_{h_2}(u_1,du_2)\dots q_{h_n}(u_{n-1},du_n)]\\
= i\sum_{n=1}^\infty i^n \int_{x_0<\dots<x_n} dx_0 \dots dx_n \ 
p(du_0) q_{h_1}(u_0,du_1) \dots q_{h_n}(u_{n-1},du_n)\\
\times [u_0 \frac{v'}{v}(x_0)+\dots +u_n \frac{v'}{v}(x_n)]\ u_0 v(x_0) \dots
u_n v(x_n) \\
= i\sum_{n=1}^\infty i^n \int_{x_0<\dots<x_n} dx_0 \dots dx_n\  u_0 v(x_0) 
\dots u_n v(x_n) \\
\times p(du_0) q_{h_1}(u_0,du_1) \dots q_{h_n}(u_{n-1},du_n)\\
\times \{ u_1 \frac{q'_{h_2}}{q_{h_2}}(u_1,du_2) +
u_2 [\frac{q'_{h_3}}{q_{h_3}}(u_2,du_3) - 
\frac{q'_{h_2}}{q_{h_2}}(u_1,du_2)]+ \dots \\
+u_{n-1} [\frac{q'_{h_n}}{q_{h_n}}(u_{n-1},du_n) 
-\frac{q'_{h_{n-1}}}{q_{h_{n-1}}}
(u_{n-2}, du_{n-1})] - u_n \frac{q'_{h_n}}{q_{h_n}} (u_{n-1},du_n) \}
\end{multline}
(by integrating by parts; we note $h_j=x_j-x_{j-1}$ and $q'_h= \partial 
q_h/\partial h$). 

This equality is equivalent to  the following infinite set of equations~:
$\forall \, n \in {\bf N^*}, \forall \, x_1<\ldots <x_n$~:

\begin{multline}
2\partial_t {\bf E}[u(x_1) \dots  u(x_n)] =\\  \int u_1\dots u_n \ 
p(du_1) q_{h_2}(u_1,du_2)\dots q_{h_n}(u_{n-1},du_n) \\
\times \{ u_1 \frac{q'_{h_2}}{q_{h_2}}(u_1,du_2) +
u_2 [\frac{q'_{h_3}}{q_{h_3}}(u_2,du_3) - 
\frac{q'_{h_2}}{q_{h_2}}(u_1,du_2)]+ \dots \\
+u_{n-1} [\frac{q'_{h_n}}{q_{h_n}}(u_{n-1},du_n) 
-\frac{q'_{h_{n-1}}}{q_{h_{n-1}}}
(u_{n-2}, du_{n-1})] - u_n \frac{q'_{h_n}}{q_{h_n}} (u_{n-1},du_n) \}
\end{multline}

One then gets the evolution equations for $p$, $q$ and $A$ by taking
limits in
which some of the $x_i$s are equal.
If one makes every $x_i$ tend to $x_1$, the preceding set of equations gives
formally, $\forall \, n \in {\bf N^*}$~: 
\begin{eqnarray*}
2\int \partial_t p(du) u^n &=&\int p(du) (-U AU^n + U^n
AU)(u)
\end{eqnarray*}  
where $U^n$ denotes the function $u\mapsto u^n$.

If one makes some of the $x_i$s tend to $x_1$, and the others tend to 
$x_2=x_1+h$, one
then gets $\forall \, n \in {\bf N^*}, \forall \, k < n, \forall \, x_1 \in 
{\bf R}, \forall \, h \in {\bf R}^{+*}$~:
\begin{multline}
2\partial_t {\bf E}[u(x_1)^k u(x_1+h)^{n-k}] =  \int p(du)\{-UA(U^k Q_h 
U^{n-k}) + U^{k+1} Q_h AU^{n-k}  \\+ U^k[A(UQ_h U^{n-k})
- AQ_h U^{n-k+1} +Q_h(U^{n-k}AU) -Q_h(U AU^{n-k})]\}(u)
\end{multline}

One then easily finds, if $\eta$ is in the domain of $A$~:
\begin{eqnarray}
2\int \partial_t p(du) \eta(u) &=&\int p(du) [-u\, A\eta (u)+
\eta (u)\, AU(u)]  \label{evp} \\
2\partial_t Q_h\eta &=& U\, AQ_h\eta + A(U \,Q_h\eta) - Q_h (U\,
A\eta) \label{evq}\\ \nonumber  
&&- A Q_h(U \,\eta) + Q_h(\eta \, AU) - AU \,Q_h\eta 
\end{eqnarray}  

These two equalities sum up into one~: $\forall \, \eta $ in the
domain of $A$,
\begin{equation}2\partial_t A \eta = U\, A^2 \eta - A^2 (U \,\eta) + A (\eta 
\,AU) -
AU\, A\eta \label{evAeta}\end{equation}
or, introducing the operators $M_U$ and $M_{AU}$ 
defined as $M_U \eta(u) = u \eta(u)$ and $M_{AU} \eta(u) = AU(u) \eta(u)$~:

\begin{equation}
2 \partial_t A = M_U A^2 - A^2 M_U + A M_{AU} - M_{AU} A
\label{evA}
\end{equation}
If this latter equality holds, one can easily check that if \ $^t\!\!A p = 0$
for all time, then $p$ verifies (\ref{evp}), and $Q_h=\exp(hA)$ verifies
(\ref{evq}). 

Hence a Feller statistical solution of (\ref{evE}) is solution of (\ref{evp})
and (\ref{evq}), which are equivalent to (\ref{evA}).

Conversely, it is a matter of simple algebra to check that (\ref{evp}) and
(\ref{evq}) imply (\ref{evE})~: 
indeed one can then write for any $x_1<\dots<x_n$  (recall $h_i =
x_i-x_{i-1}$)~:

\begin{multline}
2\partial_tE[u(x_1)\dots u(x_n)] = 2 \partial_t \int p(du) M_U Q_{h_2} \dots 
M_U Q_{h_n} U (u)\\
= 2 \int \partial_t p(du) M_U Q_{h_2} \dots M_U Q_{h_n} U (u)\\
+ 2 \sum_{j=2}^n \int  p(du) \, M_U Q_{h_2}\dots M_U Q_{h_{j-1}}  M_U \, 
\partial_t Q_{h_j} M_U Q_{h_{j+1}} \dots M_U Q_{h_n} U (u) \\
=\int p(du)  u [-AM_UQ_{h_2}M_U\dots
Q_{h_n} U (u) + Q_{h_2} M_U
\dots Q_{h_n}U (u) \, AU(u)]\\
+\sum_{j=2}^n \int p(du)\,  M_U Q_{h_2} \dots M_U [M_U A Q_{h_j} \eta_j
+A(U \, Q_{h_j} \eta_j) - Q_{h_j} (U \, A \eta_j) \\- A Q_{h_j} (U \, \eta_j) 
+ Q_{h_j} (\eta_j \, AU ) - AU \, A\eta_j ]
\end{multline}
where $\eta_j = M_U Q_{h_{j+1}} \dots M_U Q_{h_n} U$. Many terms cancel, one 
gets
\[
=\int p(du) u \sum_{j=2}^n Q_{h_2} M_U \dots Q_{h_{j-1}} M_U [M_U 
Q'_{h_j}-Q'_{h_j} M_U] \eta_j
\]
which is just one integration by parts away from (\ref{evE}).
 
Therefore, if $u(x,t)$ is a Feller process, it is a statistical solution of
Burgers if and only if its infinitesimal generator is solution of (\ref{evA}).
In some sense, the Feller assumption yields an exact closure of the infinite
set (\ref{evE}). Of course, nothing guarantees the existence of solutions of
(\ref{evA}), although we show later that the Brownian and white noise initial
cases give formal solutions to it. Nevertheless, a close look at Bertoin's 
proof using Hopf-Cole \cite{Ber} makes us strongly suspect that 
the absence of positive jumps may be essential to guarantee the
existence of solutions. This would also be reasonable from a physical
point of view : solutions with positive jumps are unphysical.

\section{The case of L\'evy processes}

We will see how one can recover formally the results of~\cite{CD}. The 
initial velocity
$u_0$ is here supposed to be a L\'evy process (which means that it has
independent and stationary increments) of finite
variance having no negative jumps. This covers in particular the case of  
$u_0$
Brownian.
 Such processes are characterized by their second
exponent $\phi $, defined by $\forall \, x<y, \forall \, \lambda \in
\RR^+$~:
\[
E\{ \exp \, [\lambda(u_0(y) - u_0(x))] \} = \exp[(y-x)\phi(\lambda)]
\]
A L\'evy process can
be considered as a limit case of stationary Markov process 
(the one point 
distribution $p$ is replaced with Lebesgue measure). One can also
formally define an infinitesimal generator by the relations~: $\forall \, 
\lambda \in \RR^+$,
\[
Ae_\lambda = \phi(\lambda) e_{\lambda}
\]
where we have noted $e_\lambda$ the function $u\mapsto \exp(\lambda u)$ (which
of course is not in the domain of $A$ \dots).
One can inject these relations into the evolution equation (\ref{evA}). 
Using $u e_\lambda(u) = \partial_\lambda e_\lambda(u)$, and $AU =$ constant,
one gets
an evolution equation for $\phi$~; it turns out that this equation is
also the Burgers equation~:
\begin{equation}
2 \partial_t \phi(\lambda) = -\partial_{\lambda}\phi^2 \label{evPhi}
\end{equation}
Carraro and Duchon~\cite{CD} have checked that if $\phi_0$ is the
exponent of a L\'evy process of finite
variance with negative jumps,  (\ref{evPhi}) has a smooth solution
for all time $t\geq 0$, which is still the exponent of a homogeneous L\'evy
process with negative jumps.

Hence such L\'evy processes are conserved by the Burgers equation. The
Brownian case corresponds to $\phi_0(\lambda) = \lambda^2/2$, and this
yields $\phi(\lambda,t) = (1+\lambda t - \sqrt{1+2\lambda t})/t^2$.

\section{Evolution equation for the jump process}

The infinitesimal generator of an arbitrary Markov process can be
written as the sum of three terms (see~\cite{Yor})~: a diffusion term,
a drift term and a jump term~: 
\[
Af(x) = a(x) f''(x) + b(x) f'(x) + \int n(x,dy) (f(y)-f(x))
\]
The measure $n(x,dy)$ represents the jump part of the process~: it gives the
number of jumps going from $x$ to $y$. In our case, all these
coefficients will of course depend on time. To write an evolution equation
for $n$ , we assume $b=1/t$ and $a=0$ for $t>0$ , and all jumps are negative.
Equation (\ref{evA})  then yields an 
evolution equation for
$n$~: $\forall \, x>y$,
\begin{multline}
  2\partial_t n(x,dy,t) = \frac{1}{t} (x-y)(\partial_x n(x,dy,t) -
\partial_y n(x,dy,t)) \\ + \int_{-\infty}^x n(x,du,t)[(x-y) n(u,dy,t) +
(y-u) n(x,dy,t)]\\ - \int_{-\infty}^y (x-u)n(y,du,t) n(x,dy,t)
\end{multline}

\section{The case of an initial white noise process}

Frachebourg and Martin~\cite{FM} have investigated the case of an initial
white noise velocity. Using the Hopf-Cole construction, they obtain
explicit formulas for the laws of $u(x,t)$ and its jumps. They actually
rederived results about Brownian motion with a parabolic drift that
had been previously established by Groeneboom~\cite{Gr} out of the
Burgers context. Using Frachebourg and Martin's results or
Groeneboom's paper, the infinitesimal generator in the case of an
initial white noise process is found to be, in the case where
$<u_0(x)u_0(y)> = (1/8)\delta(x-y)$~:
\[
Af(x) = \frac{1}{t}f'(x) + 4 \int_{-\infty}^x (f(y) - f(x)) 
(x-y)\frac{J(yt^{1/3})}{J(xt^{1/3})}
I(xt^{1/3}-yt^{1/3}) dy
\]
where $I$ and $J$ are given by their Fourier and Laplace transforms in
terms of the Airy function Ai~:
\begin{eqnarray}
J(u)&=&\frac{1}{2i\pi} \int_{-i\infty}^{i\infty} dz
\frac{\exp(uz)}{2^{1/3}\mbox{Ai}(2^{-1/3}z)} \label{FourJ}\\
2I(u)&=& (2\pi u^3)^{-1/2} +
\frac{1}{2i\pi}\int_{-i\infty}^{i\infty}\exp(uz)(\frac{2^{2/3}
  \mbox{Ai}'(2^{-1/3} z)}{\mbox{Ai}(2^{-1/3}z)} +(2z)^{1/2})\label{LaplI}
\end{eqnarray}
    
We have checked that the evolution equation (\ref{evA}) is indeed verified~: 
it amounts to
expressing convolutions like $uI\ast J$, $uI\ast uI$, $uI\ast uJ$
in terms of $J'$ and $I'$. It can be done using relations (\ref{FourJ}) and
(\ref{LaplI}) and the fact that $\mbox{Ai}''(x) = x \mbox{Ai}(x)$.

\section{Conclusion}

We have heuristically shown that for Feller stationary processes, Burgers
equation is equivalent to an evolution equation for their infinitesimal 
generators. It gives strong evidence that the Feller property is conserved
by Burgers equation, although we suspect that the negativity of jumps
in the initial velocity
should be required.
Our evolution equation provides an equation for the
jump process, and this might lead to other exact statistical solutions of
Burgers equation. The previous exact solutions concerning an initial Brownian
 or white noise velocity are both particular solutions of our equation.

\bibliographystyle{plain}

\end{document}